\documentclass[manuscript,screen]{acmart}
\usepackage{tabularx}

\def\BibTeX{{\rm B\kern-.05em{\sc i\kern-.025em b}\kern-.08em
    T\kern-.1667em\lower.7ex\hbox{E}\kern-.125emX}}
\usepackage{amsmath}  
\usepackage{graphicx} 
\usepackage{tikz}
\usetikzlibrary{shapes.geometric, arrows, positioning}
\usepackage{pgfplots}
\pgfplotsset{compat=1.18}
\usepackage{subcaption}
\usepackage{wrapfig}

\AtBeginDocument{%
  \providecommand\BibTeX{{%
    Bib\TeX}}}

\setcopyright{acmlicensed}
\copyrightyear{2024}
\acmYear{2024}
\acmDOI{}

\begin{document}

\title{Modeling and Simulation of a Fully Autonomous Electric Vehicle (AEV)}

\author{Qasim Ajao}
\email{qasimajao@ieee.org}
\orcid{0009-0007-7371-1742}
\author{Lanre Sadeeq}
\authornotemark[]
\email{}
\orcid{0009-0000-3075-8339}
\affiliation{
  \institution{National Institute of Technology}
  \city{Lagos}
  \country{Nigeria}
}



\begin{abstract}
With the continuous advancements in science and technology, there is an increasing focus on environmental sustainability, leading to heightened interest in \textbf{autonomous electric vehicles (AEVs)}. AEVs hold significant potential for enhancing electric mobility, energy efficiency, environmental preservation, and driving capabilities. They offer numerous advantages that could revolutionize transportation and urban lifestyles, notably improving road safety by reducing human error—a leading cause of accidents. This paper presents a comprehensive simulation of an AEV by developing and integrating models for the driving system, battery, motor, transmission, and vehicle body within the MATLAB/Simulink environment. Each component is configured and interconnected to create a pure vehicle model. The simulation is performed under UDDS cycle conditions to evaluate the vehicle’s performance in \textbf{speed maintenance, driving distance, acceleration capabilities, and battery state-of-charge (SoC) dynamics}. The results demonstrate that the car exhibits strong output characteristics. Furthermore, the driver model incorporates an energy brake recovery function, which, when set to a 50\% recovery efficiency, increases the driving distance by 25\% compared to a vehicle without this function.
\end{abstract}





\maketitle

\section{INTRODUCTION}
While traditional \textbf{internal combustion engine (ICE)} vehicle technology has advanced considerably, the rapid increase in vehicle numbers has led to heightened fossil fuel consumption and significant environmental concerns. This has accelerated the need for alternative automotive technologies that do not rely on conventional energy sources \cite{b1}. \textbf{Autonomous electric vehicles (AEVs)}, known for their energy efficiency, sustainability, and superior driving capabilities, have emerged as a key focus for the 21st-century automotive industry \cite{b2}. Research into AEVs has expanded across various dimensions, particularly in enhancing their autonomous capabilities and integrating advanced electric powertrains \cite{b3,b4}. For instance, to address the impact of drivetrain flexibility in AEVs, researchers have developed electric drive systems that incorporate a double-inertia model augmented with notch filters to improve transmission accuracy \cite{b5}. Additionally, efforts to optimize the powertrain parameters of AEVs under performance constraints have led to driving range improvements and power consumption reductions \cite{b6}. The power battery, often considered the “core” of an AEV, has also been a major research focus \cite{b7}. Predictive models for battery state-of-charge (SoC) that use advanced algorithms like decision tree regression and light gradient boosting have shown to be very accurate, enhancing battery management for self-driving operations \cite{b8}. Moreover, improved modeling techniques have addressed SoC estimation challenges influenced by factors like charge and discharge rates and temperature \cite{b1}. Enhancements to the refined Kalman filter parameters have made SoC estimation more accurate, which is crucial for AEV dependability. As the demand for AEVs grows, research into their charging infrastructure has also intensified. Studies have developed optimization models that balance power grid needs and autonomous operation requirements, using sophisticated algorithms to manage peak loads and ensure continuous operation \cite{b6}. Additionally, researchers have integrated models for autonomous charging and battery swapping systems to create dual-layer energy exchange frameworks, effectively supporting the unique needs of AEVs \cite{b9}.

Despite these advancements, comprehensive research on the modeling of autonomous electric vehicles (AEVs) remains relatively limited, yet it is essential for their successful development. This paper seeks to bridge this gap by establishing detailed models for the autonomous driving system, battery, motor, transmission, and vehicle body within the MATLAB/Simulink environment. The model also includes an energy brake recovery system designed specifically for AEVs. We organize the remainder of this paper as follows: Section II presents the literature review; Section III provides a design of the proposed vehicle model; Section IV details the specifications, parameters, and nomenclature used in the paper; Section V presents the simulation results; and Section VI concludes this paper.

\section{PREVIOUS WORK}
The growing demand for sustainable transportation has driven extensive research into the modeling and simulation of electric vehicles (EVs) \cite{b1}, which are crucial for enhancing performance, efficiency, and reliability. This chapter reviews significant advancements in this field, highlighting key studies that have contributed to the evolution of EV technology. Battery technology remains a critical factor in EV performance. Researchers have developed sophisticated battery models, such as an improved equivalent circuit model using Modelica, which simulates internal battery polarization with high accuracy \cite{b10}. Comprehensive simulations have validated these models, establishing them as valuable tools for multi-domain system modeling in EVs. The performance of EVs is also influenced by powertrain and vehicle dynamics. Studies using tools like ADVISOR have modeled and simulated various vehicle topologies, providing insights into performance metrics, energy consumption, and emission profiles. These dynamic models are essential for accurately assessing and optimizing EV performance \cite{b11}.

Validation through simulation tools is critical for ensuring high performance \cite{b12}. MATLAB has been highlighted as a robust platform for modeling, validation, and simulation of EV powertrains. Research has shown MATLAB’s capabilities in predicting real-world performance and aiding decision-making in EV design and management \cite{b13}. Studies using MATLAB/Simulink have demonstrated superior speed-following performance, extended driving range, and reduced battery SOC, with energy brake recovery systems further improving efficiency \cite{b14}. Previous research has also focused on detailed modeling and simulation of EVs under various driving conditions, such as the New European Driving Cycle (NEDC). These studies have shown superior speed-following capabilities and extended driving ranges, emphasizing the importance of advanced modeling techniques in optimizing EV performance and meeting consumer needs \cite{b1,b15}.

While EV modeling and simulation have seen substantial progress, the integration of fully autonomous driving technologies presents the next significant challenge for researchers and scholars in this field. Autonomous electric vehicles (AEVs) hold the potential to revolutionize transportation by enhancing safety, reducing traffic congestion, and increasing energy efficiency \cite{b16}. However, integrating these technologies requires advanced modeling and simulation approaches capable of addressing the complexities of autonomous operation. Developing and validating sophisticated algorithms for navigation, obstacle detection, and real-time decision-making in simulated environments is essential before their safe implementation in real-world scenarios \cite{b3,b7,b9}.

\section{DEVELOPMENT OF THE AEV MODEL}
The design of the autonomous electric vehicle (AEV) model incorporates cutting-edge technologies to create a highly efficient and autonomous operation. The central component of a generic AEV design is a sophisticated sensor suite, comprising LiDAR, radar, and high-resolution cameras, that enables a comprehensive understanding of the surrounding environment. The car’s powertrain prioritizes electric propulsion to maximize energy economy and sustainability. Additionally, AEV's model includes resilient communication systems for vehicle-to-vehicle (V2V) and vehicle-to-infrastructure (V2I) interactions, greatly improving safety and traffic coordination. The integration and performance of these components undergo rigorous validation using a combination of detailed simulations and real-world testing to ensure reliability and robustness under various driving circumstances \cite{b1,b2,b3,b9,b12}. In this paper, the driving dynamics, battery configuration, motor model, transmission design, and body model of an AEV in Simulink are established, connecting and combining all components to form a pure vehicle model.

\subsection{\textbf{Driver Model}}
The simulation detailed in this paper extensively examines the vehicle's dynamic behavior, with particular emphasis on its longitudinal dynamics. Consequently, we specifically design the driver model to focus on the vehicle's longitudinal dynamics, accurately simulating throttle and braking actions. The objective is to align the vehicle's current speed with the target speed, which is achieved through the use of proportional-integral (PI) control. Therefore, the fundamental equation that guides this control is as follows:

\begin{equation*}
v = m_{\mathrm{a}} \times \Delta v + m_{\mathrm{b}} \times \int \Delta v \, d t \tag{1}
\end{equation*}

\begin{figure}[ht]
\centering
\includegraphics[width=1\textwidth]{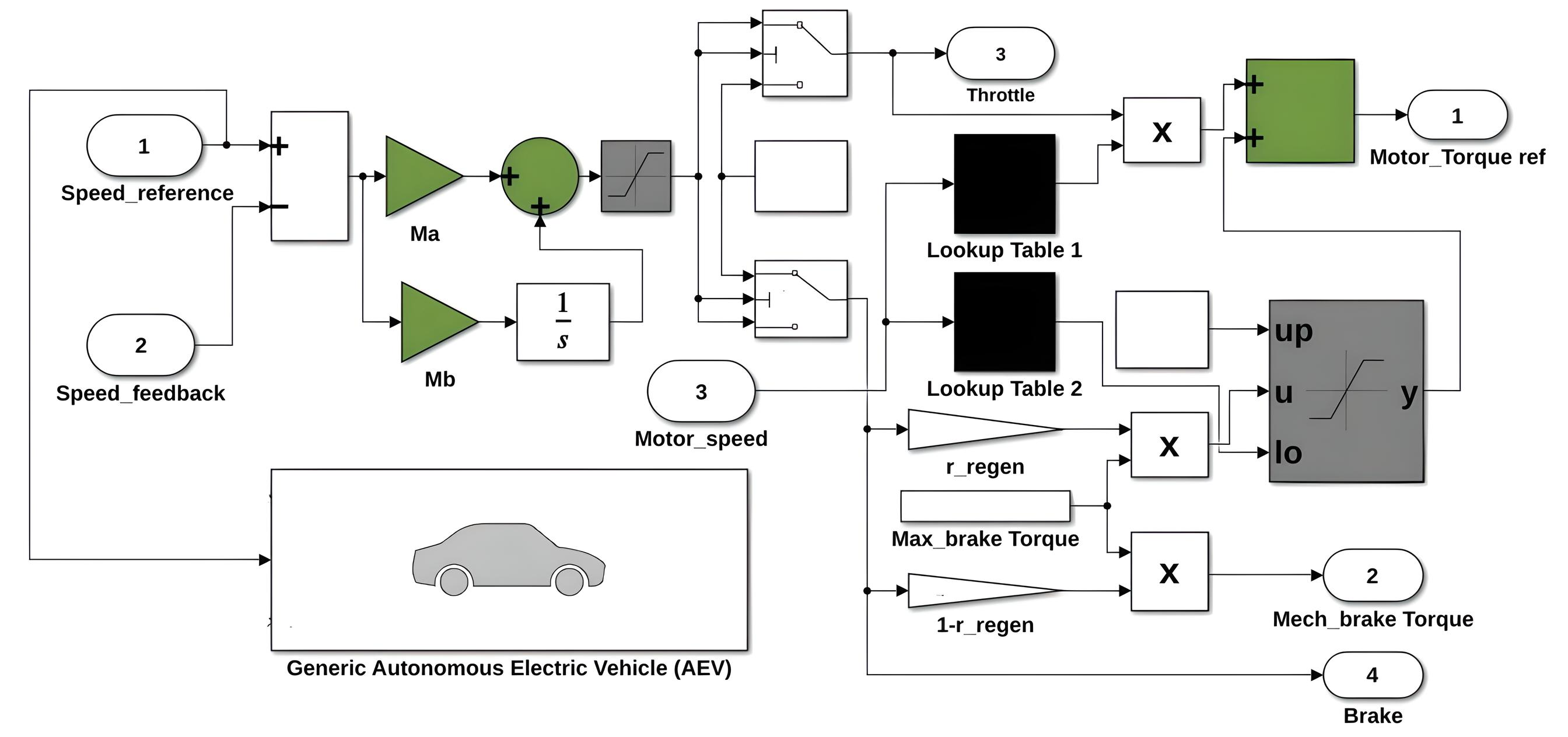}
\caption{Driver Model}
\Description{Driver Model}
\end{figure}

where \( \Delta v \) represents the difference between the expected speed and the actual speed, and \( m_{\mathrm{a}} \) and \( m_{\mathrm{b}} \) are the coefficients for proportional and integral gains, respectively. This is processed by the PI controller to produce an output ranging from -1 to 1. A positive output, between 0 and 1, indicates that the current speed is below the target speed, prompting the driver to accelerate. Conversely, a negative output, between -1 and 0, suggests that the current speed exceeds the target speed, leading the driver to apply the brakes to achieve the desired speed. Additionally, the model incorporates a regenerative braking mechanism to enhance energy efficiency. The braking process converts the typically lost mechanical energy into electrical energy, thereby recharging the battery. Figure 1 depicts the developed driver model.

\subsection{\textbf{Battery Model}}
The battery serves as the primary energy source for autonomous electric vehicles, converting stored electrical energy into the vehicle's kinetic energy. The battery model is constructed using the ampere-hour (Ah) counting method, with its fundamental principle defined as follows:

\begin{equation*}
S O C_k = S O C_0 - \frac{1}{C_b} \int_{t_0}^{t_k} \eta J \, dt \tag{2}
\end{equation*}

where \(J\) represents the current passing through the battery, and the positive values signify discharging, and negative values indicate a charging state during regenerative braking. The term \(C_b\) refers to the battery’s total capacity, while \(S O C_0\) and \(S O C_k\) indicate the initial and current state of charge, respectively. Besides calculating the state of charge (SoC), this model also predicts the battery’s voltage and power. The subsequent equation determines the voltage by subtracting the voltage drop, a result of internal resistance, from the nominal voltage. 

\begin{equation*}
V = V_{\mathrm{n}} - Z J \tag{3}
\end{equation*}

The corresponding expression for power is:

\begin{equation*}
P = \int_{t_{\mathrm{o}}}^{t_{\mathrm{k}}} V J \, dt \tag{4}
\end{equation*}

where \( V \) represents the voltage, \( V_{\mathrm{n}} \) is the nominal voltage, \( Z \) stands for the internal resistance, \( P \) denotes the power, \( J \) is the current, and \( t \) is the time variable.

\begin{figure}[ht]
\centering
\includegraphics[width=0.9\textwidth]{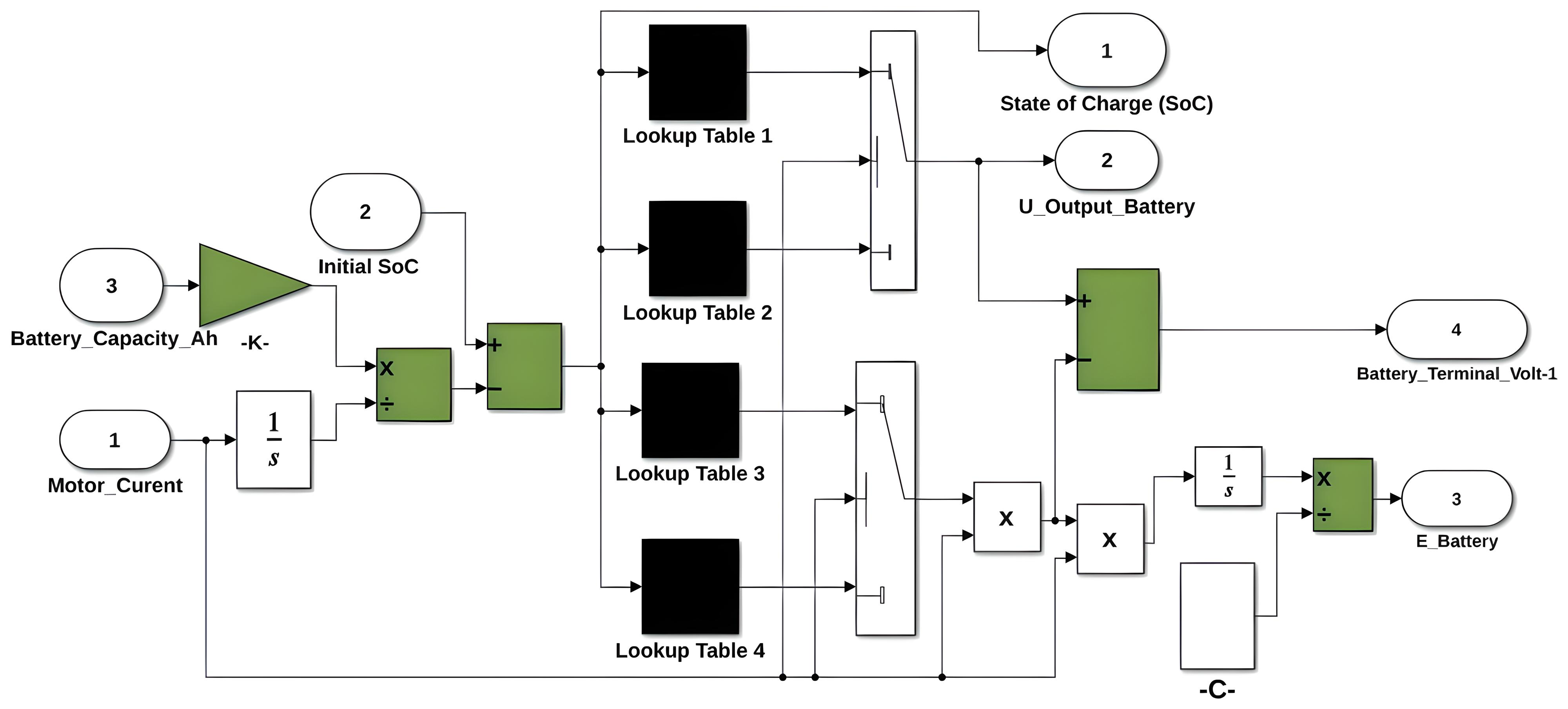}
\caption{Battery Model}
\Description{Battery Model}
\end{figure}

\subsection{\textbf{Motor Model}}
The motor model calculates the motor's required current using inputs such as motor torque, rotational speed, and battery voltage. This current represents the electrical flow back into the battery. The following formula provides the motor power expression:

\begin{equation*}
P_{\mathrm{motor}}=\frac{\tau \times n \times \eta}{9550} \tag{5}
\end{equation*}

In this equation, \(P_{\mathrm{motor}}\), \(\tau\), \(n\), and \(\eta\) correspond to the motor’s power output, torque, speed, and efficiency, respectively. Following this, the motor current can be determined based on the relationship between power, voltage, and current using the formula:

\begin{equation*}
J_{\mathrm{motor}}=\frac{P}{V} \tag{6}
\end{equation*}

\begin{figure}[ht]
\centering
\includegraphics[width=0.9\textwidth]{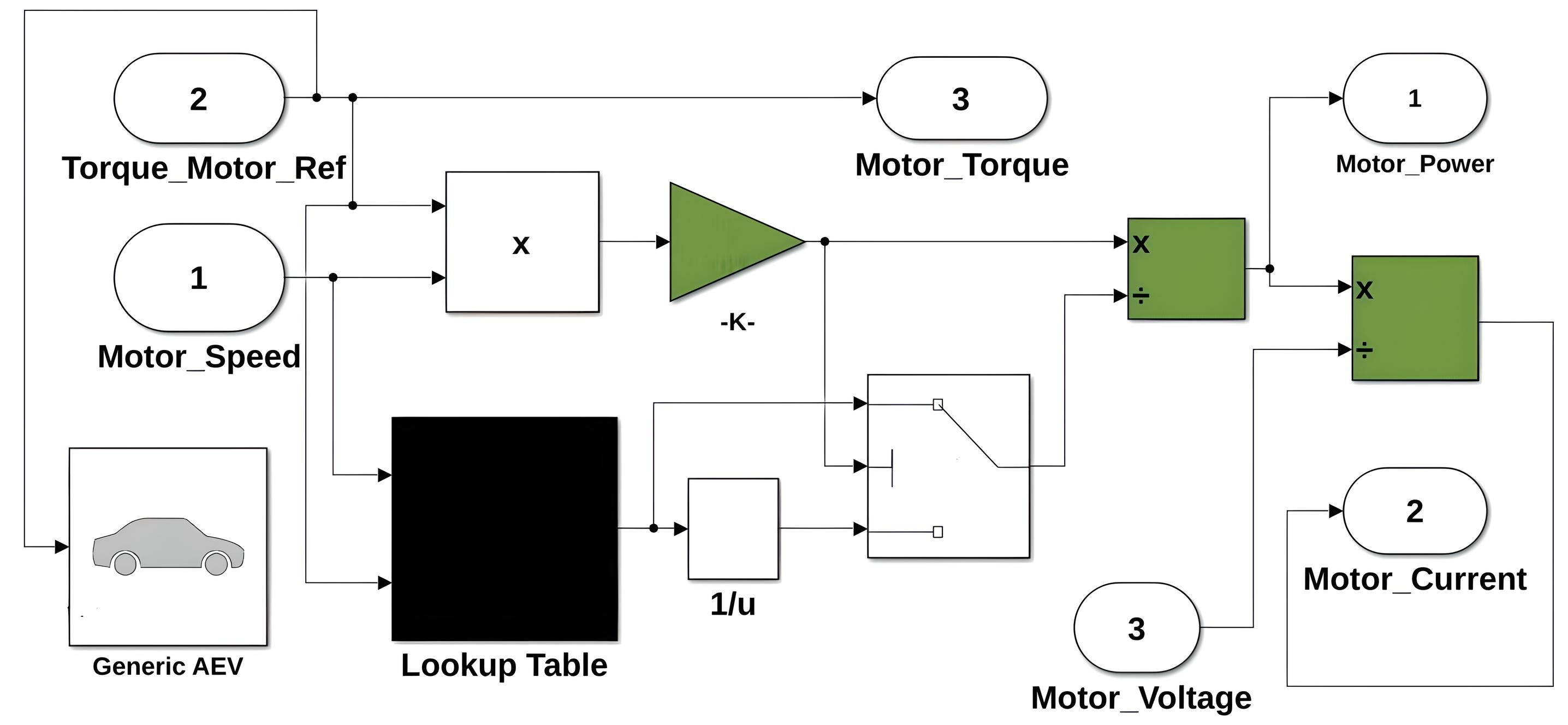}
\caption{Motor Model}
\Description{Motor Model}
\end{figure}

\subsection{\textbf{Transmission Dynamics Model}}
The control model transmits the propulsion torque to the wheels through the drivetrain, enhancing the AEV's acceleration or deceleration and optimizing its overall dynamic performance. The core equation governing the transmission dynamics model is as follows:

\begin{equation*}
\tau = M_{\mathrm{\tau}} \times G_{\mathrm{R}} \times \eta \tag{7}
\end{equation*}

where \(M_{\mathrm{\tau}}\) represents the torque generated by the control model, \(G_{\mathrm{R}}\) denotes the gear ratio, and \(\eta\) indicates the efficiency of the transmission system. The designed transmission dynamics model is illustrated in Figure 4 below:

\begin{figure}[ht]
\centering
\includegraphics[width=0.9\textwidth]{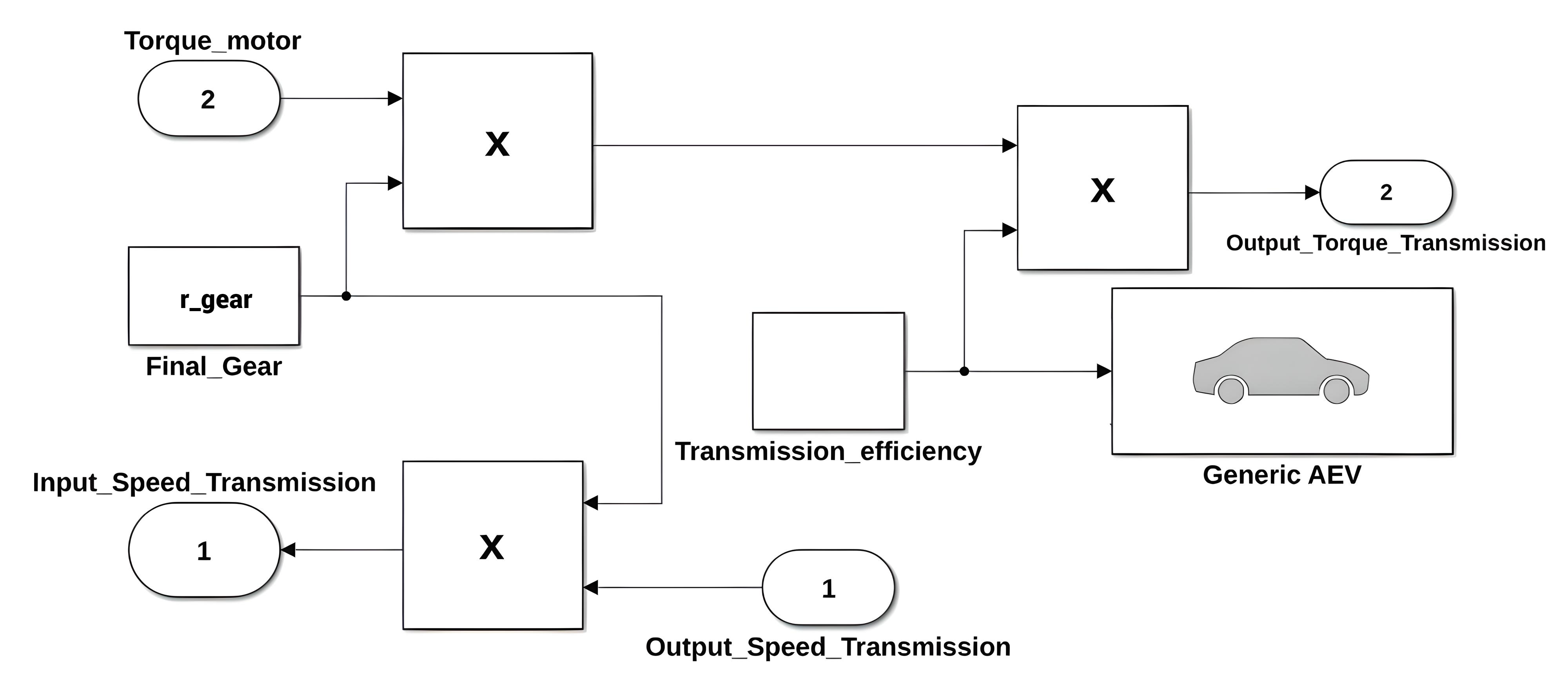}
\caption{Transmission Model}
\Description{Transmission Model}
\end{figure}

\subsection{\textbf{Body Model}}
The vehicle dynamics model processes the propulsion torque and braking torque as input parameters and computes the resulting wheel speed, distance traveled, and velocity of the AEV.

\begin{figure}[ht]
\centering
\includegraphics[width=0.9\textwidth]{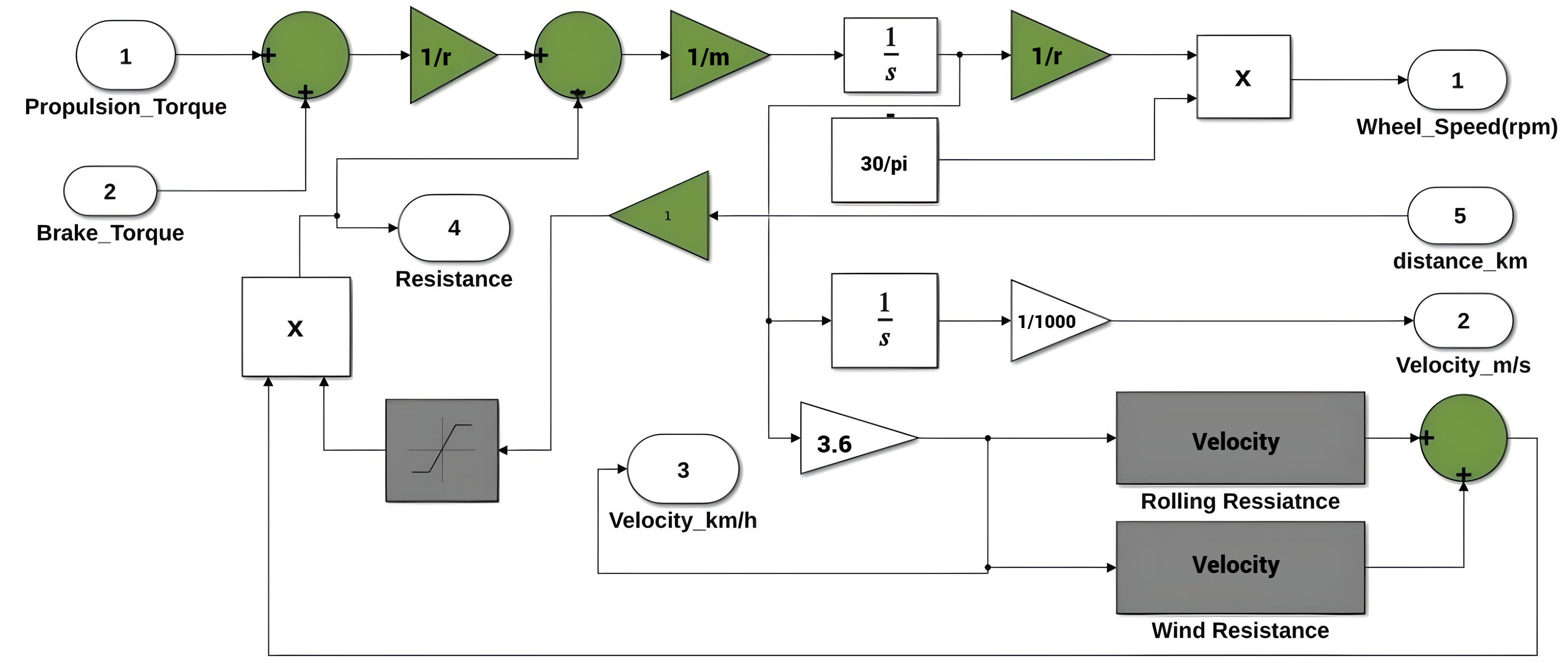}
\caption{Body Model}
\Description{Body Model}
\end{figure}

\begin{equation*}
\frac{\frac{\tau_{\mathrm{p}}+\tau_{\mathrm{b}}}{R_w}-R R-W R}{m}=a \tag{8}
\end{equation*}

In this equation, \(\mathrm{\tau}_{\mathrm{p}}, \mathrm{\tau}_{\mathrm{B}}, \mathrm{R_w}, \mathrm{m}\), and \(a\) represent the propulsion torque, braking torque, wheel radius, vehicle mass, and acceleration, respectively. The terms RR and WR correspond to the rolling resistance and aerodynamic drag (wind resistance) experienced by the vehicle during operation. We calculate the rolling resistance using the empirical formula:

\begin{equation*}
R R=m g \times\left(f_{0}+f_{1} \times \frac{n}{100}+f_{4} \times\left(\frac{n}{100}\right)^{4}\right) \tag{9}
\end{equation*}

where \(f_{0}, f_{1}\), and \(f_{4}\) are empirical coefficients, while \(g\) and \(n\) denote the gravitational acceleration and vehicle speed, respectively. 

The aerodynamic drag is determined by the following expression:

\begin{equation*}
W R=\frac{C_{\mathrm{d}} \times A_f \times n^{2}}{21.15} \tag{10}
\end{equation*}

where \(C_{d}\) is the drag coefficient, and \(A_f\) is the frontal area of the vehicle. The developed vehicle dynamics model is illustrated in Figure 5:

\subsection{\textbf{Full Vehicle Model}}
The models interconnect their outputs and inputs, utilizing the output from one subsystem as the input for another. This integration process resulted in the creation of a comprehensive autonomous electric vehicle (AEV) model. Figure 6 depicts the assembled vehicle model.

\begin{figure}[ht]
\centering
\includegraphics[width=1\textwidth]{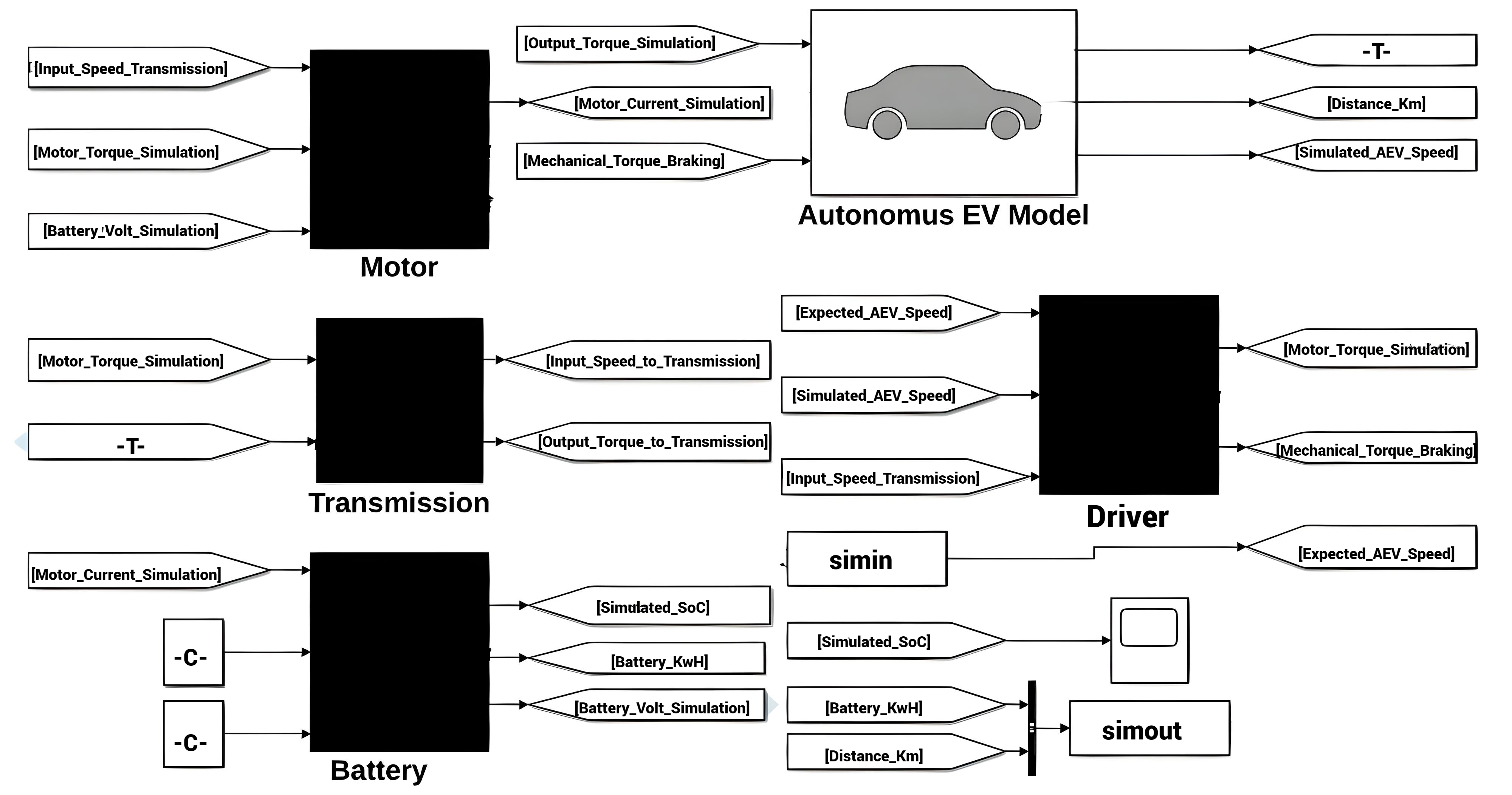}
\caption{Pure Autonomous EV Model}
\Description{Pure AEV Model}
\end{figure}

\section{PARAMETER SETTINGS}
\subsection{\textbf{Core Specifications}}
The paper utilizes the symbols and descriptions listed in Table 1. Table 2 outlines the critical specifications derived from a specific autonomous vehicle's design criteria. As shown in Equation (9), an increase in frontal area or aerodynamic drag coefficient leads to higher resistance during vehicle operation, with wind resistance being directly proportional to the square of the vehicle's velocity. The rolling resistance coefficient, denoted as $f_{0}$, is presented above, while the coefficients $f_{1}$ and $f_{4}$ are set to zero in this paper. Transmission efficiency refers to the proportion of actual transmitted power relative to the input power within the gear system. The maximum braking force indicates the peak force exerted by the vehicle's braking system. Regenerative braking efficiency is defined as the ratio of electrical energy recovered by the battery to the mechanical energy dissipated during braking.

\begin{table}[ht]
\centering
\caption{Symbols and Descriptions}
\begin{tabularx}{\columnwidth}{X l X l}
\toprule
\textbf{Symbol} & \textbf{Description} & \textbf{Symbol} & \textbf{Description} \\
\midrule
AEV & Autonomous Electric Vehicle & RR & Rolling resistance \\
UDDS & Urban Dynamometer Driving Schedule & WR & Aerodynamic drag \\
\( v \) & Vehicle speed & \( C_{\mathrm{d}} \) & Drag coefficient \\
\( m_{\mathrm{a}} \) & Proportional gain coefficient & \( A_f \) & Frontal area of the vehicle \\
\( m_{\mathrm{b}} \) & Integral gain coefficient & \( g \) & Gravitational acceleration \\
\( \Delta v \) & Speed difference (expected vs actual) & \( f_{0}, f_{1}, f_{4} \) & Empirical coefficients for rolling resistance \\
\( S O C_k \) & State of charge at time \( t_k \) & \( J \) & Current through the battery \\
\( S O C_0 \) & Initial state of charge & \( C_b \) & Battery capacity \\
\( V \) & Voltage & \( V_{\mathrm{n}} \) & Nominal voltage \\
\( Z \) & Internal resistance & \( P_{\mathrm{motor}} \) & Motor power \\
\( \tau \) & Motor torque & \( n \) & Motor speed \\
\( \eta \) & Efficiency of the system & \( G_{\mathrm{R}} \) & Gear ratio \\
\( M_{\mathrm{\tau}} \) & Torque from control model & \( R_w \) & Wheel radius \\
\bottomrule
\end{tabularx}
\end{table}

\begin{table}[ht]
\centering
\caption{BASIC VEHICLE SPECIFICATIONS}
\begin{tabularx}{\columnwidth}{X X}
\toprule
\textbf{Specification} & \textbf{Values} \\
\midrule
Vehicle Mass & 1549 kg \\
Wheel Radius & 0.284 m \\
Frontal Area & $1.87 \mathrm{~m}^{2}$ \\
Aerodynamic Drag Coefficient & 0.42 \\
Rolling Resistance Coefficient & 0.021 \\
Transmission Efficiency & 0.9 \\
Maximum Braking Force & 800 N \\
Regenerative Braking Efficiency & 0.5 \\
\bottomrule
\end{tabularx}
\end{table}

\subsection{\textbf{Motor Configuration}}
The configuration of the motor in an autonomous electric vehicle (AEV) is critical to its overall performance, particularly in terms of power delivery and responsiveness. The motor power is:

\begin{equation*}
P=\frac{V}{3600}(R R+W R) \tag{11}
\end{equation*}

Differences in motor settings can greatly impact the AEV's acceleration capabilities, top speed, and overall driving efficiency. Therefore, it is vital to carefully optimize the motor parameters to ensure that the AEV achieves the desired performance outcomes. Table 3 below displays the parameters selected for the motor in this paper, demonstrating a strategic approach to balance power, efficiency, and driving dynamics for optimal AEV operation. Integrating the parameters, the power output is calculated to be $P = 29.48 \, \text{kW}$. Based on this, a motor with a rated power of 30 kW is selected. Considering that the typical motor overload factor ranges between 2 and 3, the maximum power capacity is determined to be 75 kW. For optimal motor design, selecting a motor with a higher rated speed for the same rated power results in a more compact, lighter, and cost-effective solution. Additionally, to maintain constant motor power, an increase in rated and maximum speed will proportionally reduce the motor's maximum torque, mitigating potential issues associated with excessive torque. Therefore, we choose a high-speed motor with a rated speed of 3000 rpm and a maximum speed of 8000 rpm. According to Equation (5), the torque $\tau$ is calculated to be 95.5 Nm. Since the maximum torque typically ranges from 1.6 to 2.5 times the rated torque, we set the maximum torque for the selected motor at 230 Nm.

\begin{table}[ht]
\centering
\caption{MOTOR PARAMETERS}
\begin{tabularx}{\columnwidth}{X X}
\toprule
Parameter Name & Parameter Values \\
\midrule
Rated torque & $95.5 \mathrm{~N} \cdot \mathrm{m}$ \\
Maximum torque & $230 \mathrm{~N} \cdot \mathrm{m}$ \\
Rated power & 30 kW \\
Maximum power & 75 kW \\
Rated rotating speed & $3000 \mathrm{r} / \mathrm{min}$ \\
Maximum motor speed & $8000 \mathrm{r} / \mathrm{min}$ \\
\bottomrule
\end{tabularx}
\end{table}

\subsection{\textbf{Battery Configuration}}
The capacity of the battery is a critical factor that directly impacts the driving range of autonomous electric vehicles (AEVs). While increasing battery capacity can extend the vehicle's range, it also leads to a higher overall vehicle mass, which, in turn, increases energy consumption. Additionally, as battery capacity grows, it requires a greater number of individual cells, which complicates the battery thermal management and battery management systems (BMS), ultimately driving up costs and complexity. This trade-off suggests that simply increasing battery capacity may not always be the most efficient solution. In this paper, a battery pack with a capacity of $216 \, \text{kWh}$, commonly used in autonomous vehicles, is selected to balance range extension with manageable system complexity and cost.

\section{Simulation and Evaluation}
The developed AEV model was subjected to comprehensive simulation testing within the MATLAB/Simulink environment by inputting various vehicle parameters. The model's performance was evaluated under standard driving conditions using the UDDS.

\begin{figure}[ht]
\centering
\includegraphics[width=0.7\textwidth]{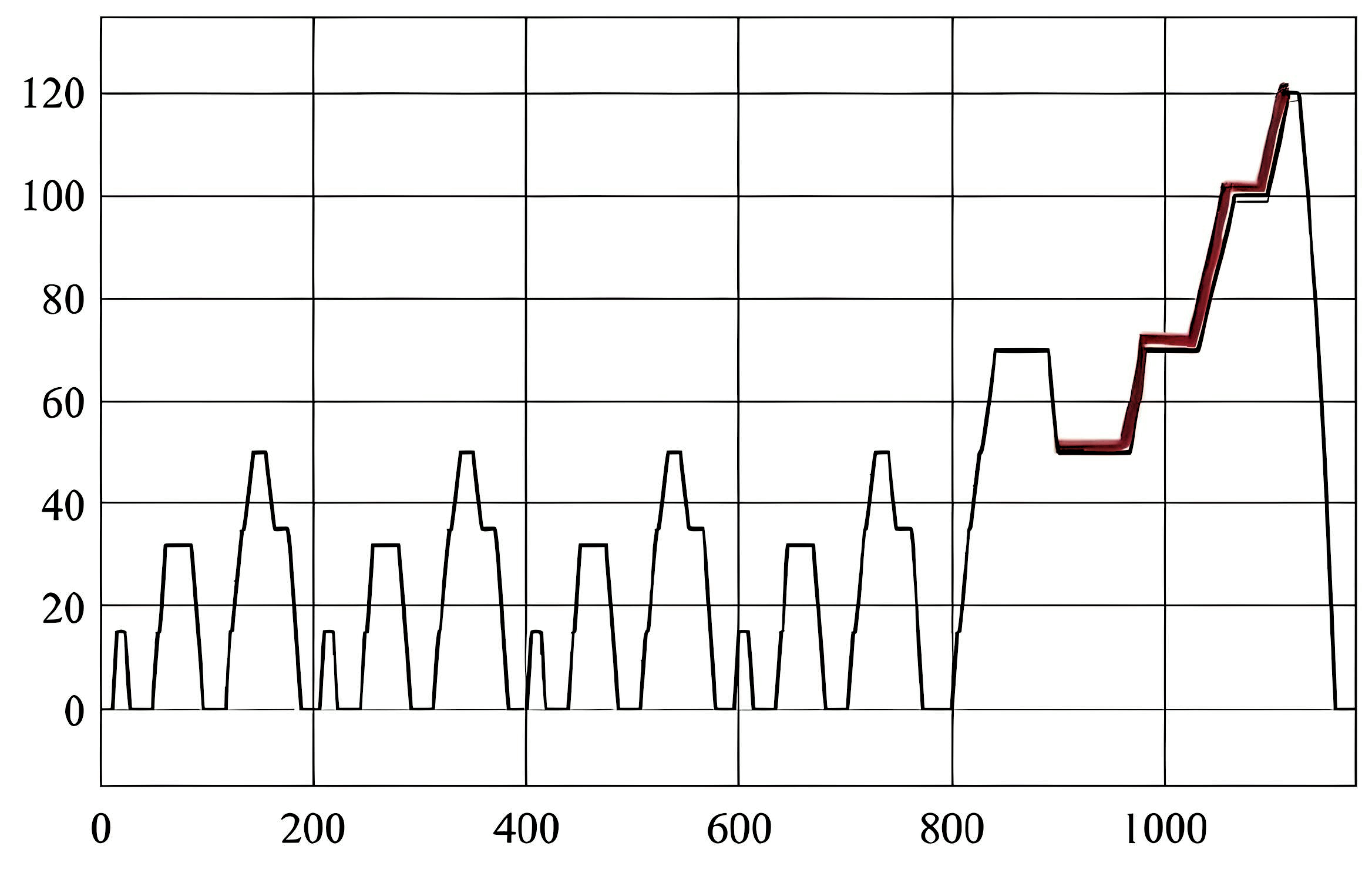}
\caption{Velocity Tracking Curve}
\Description{Velocity Tracking Curve}
\end{figure}

\subsection{\textbf{Speed/Velocity Performance}}
Velocity or speed tracking performance refers to the AEV's ability to keep its actual speed in close alignment with the desired speed. The smaller the deviation between these two, the better the tracking performance. Figure 7 illustrates the speed tracking results. In the graph above, the red curve represents the desired speed, while the black curve depicts the actual speed. The results demonstrate strong tracking performance, with the actual speed closely matching the desired speed at low velocities and only minor deviations at higher speeds, where the maximum speed error does not exceed $1.5\%$. The AEV's performance in acceleration and deceleration phases is similarly well-matched.

\begin{figure}[ht]
\centering
\begin{minipage}[b]{0.48\textwidth}
    \centering
    \includegraphics[width=1\textwidth]{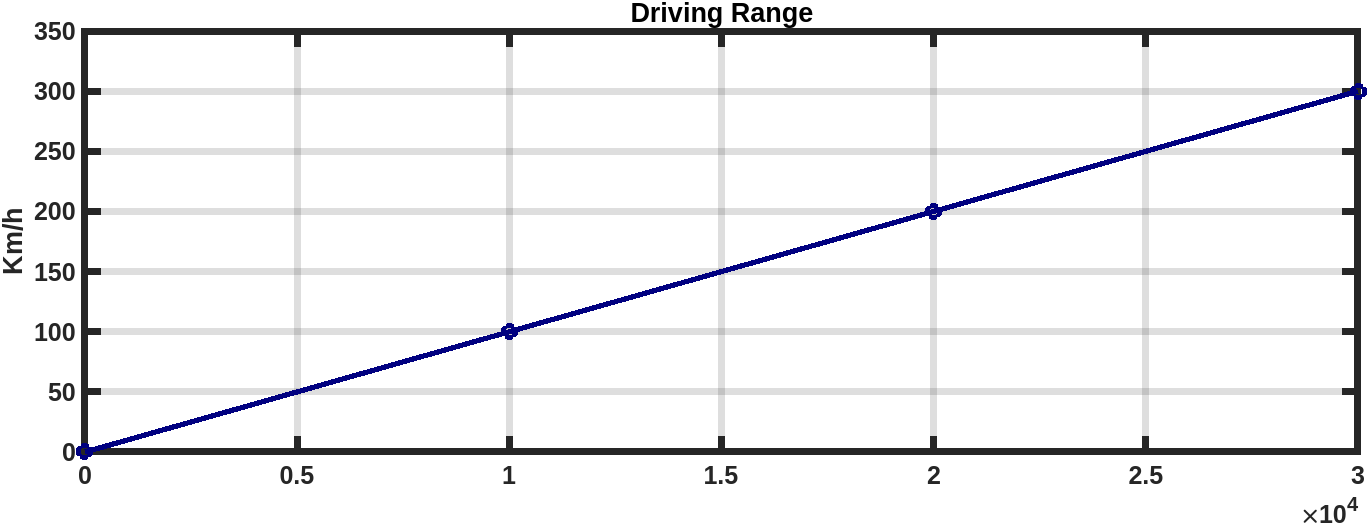}
\end{minipage}
\hfill
\begin{minipage}[b]{0.48\textwidth}
    \centering
    \includegraphics[width=1\textwidth]{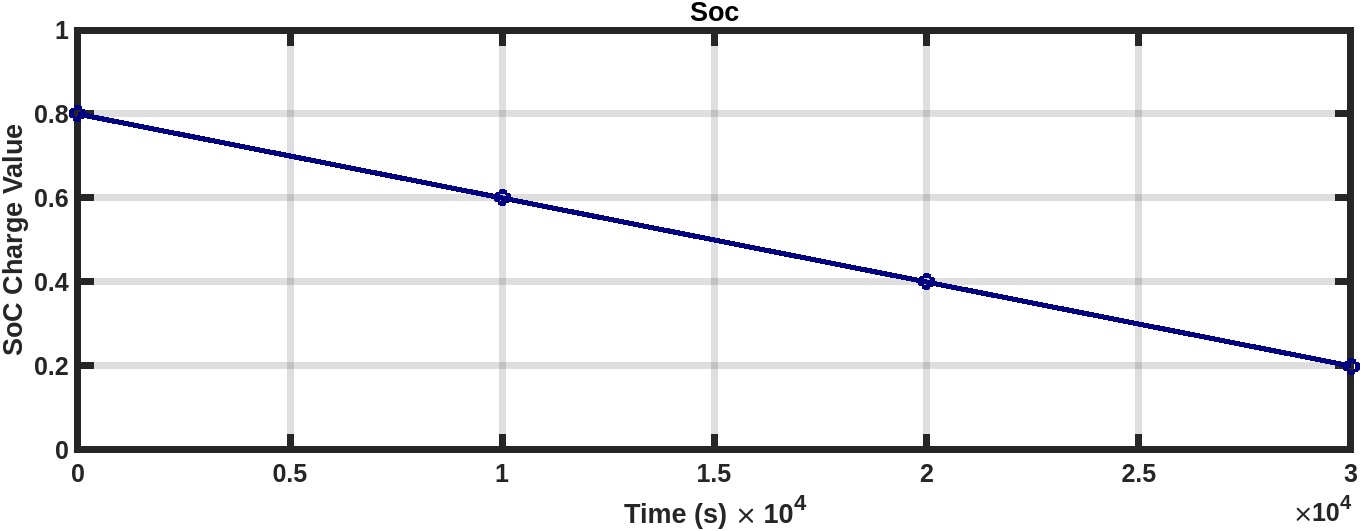}
\end{minipage}
\caption{Travel Distance and Corresponding SoC}
\Description{Travel Distance and Corresponding SoC}
\end{figure}

\subsection{\textbf{Range Analysis}}
The battery's initial State of Charge (SoC) was set to 0.9, enabling the AEV to continuously complete the UDDS cycle, from which the total driving distance was calculated. In Figure 8, the upper section shows the driving range, while the lower section tracks the corresponding SoC of the battery. The vehicle completed 32 UDDS cycles, covering approximately 352 kilometers, reducing the SoC from 0.9 to about 0.1. This demonstrates that the AEV provides sufficient range for typical daily use. For comparison, Figures 9 illustrate the driving range without the energy recovery braking system. The scenario saw a decrease in the SoC from 0.9 to about 0.1, resulting in a limited driving range of approximately 286 kilometers. This highlights the significant contribution of the energy recovery braking system, which extended the driving range by $23\%$ when operating at a recovery efficiency of 0.5.

\begin{figure}[ht]
\centering
\begin{minipage}[b]{0.48\textwidth}
    \centering
    \includegraphics[width=1\textwidth]{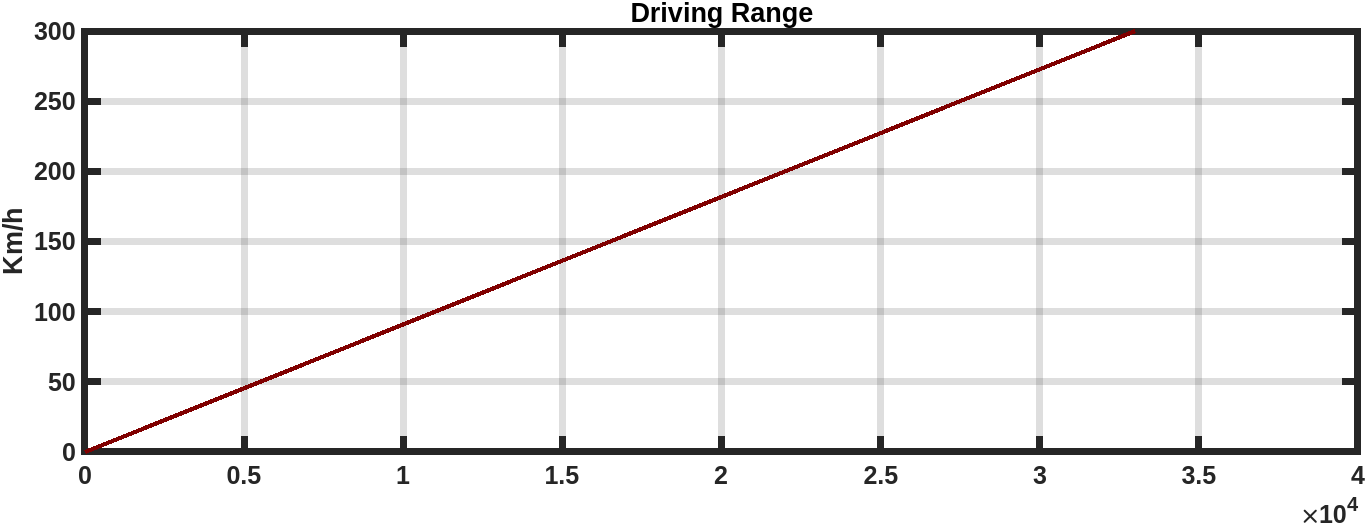}
\end{minipage}
\hfill
\begin{minipage}[b]{0.48\textwidth}
    \centering
    \includegraphics[width=1\textwidth]{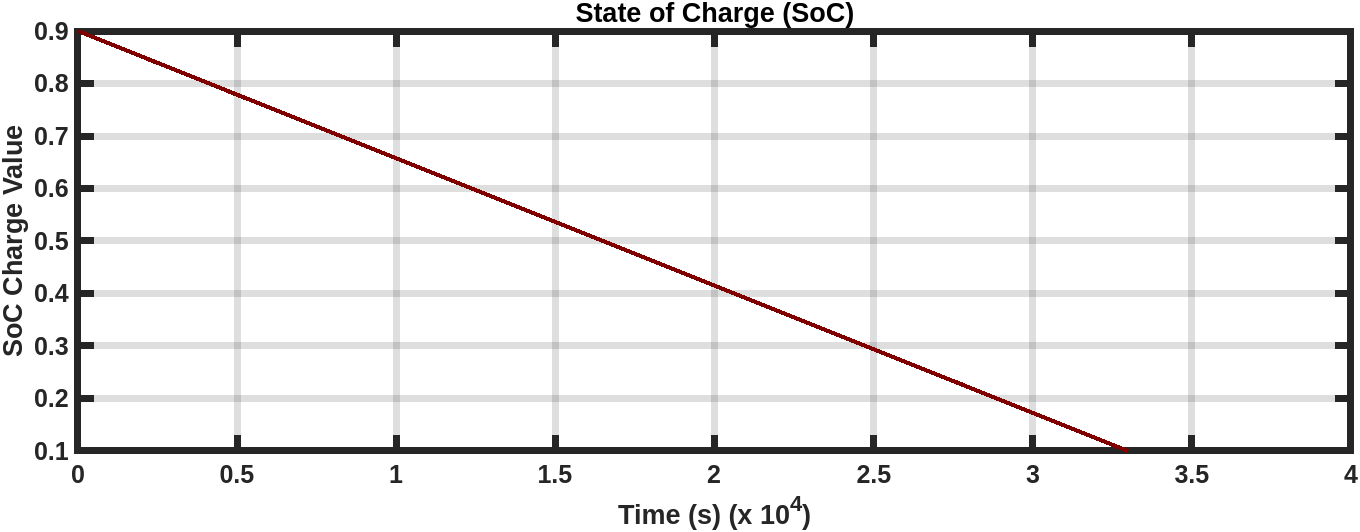}
\end{minipage}
\caption{Range curve without Energy Recovery Braking System and Corresponding SoC}
\Description{Range curve without Energy Recovery Braking System and Corresponding SoC}
\end{figure}

\subsection{\textbf{Acceleration Performance}}
Acceleration performance, particularly the time required to reach 100 km/h, is a key metric for evaluating an autonomous EV powertrain's efficiency. The vehicle's acceleration curve, aiming for a target speed of 100 km/h, is shown in Figure 10. As depicted, the blue curve indicates the target speed, while the red curve shows the actual acceleration. The vehicle accelerates in a near-linear fashion up to 90 km/h, maintaining a consistent acceleration rate. Beyond 90 km/h, the rate of acceleration gradually decreases. The AEV reaches 100 km/h in approximately 9.5 seconds.

\begin{figure}[htbp]
\centering
\begin{minipage}[b]{0.48\textwidth}
    \centering
    \includegraphics[width=1\textwidth]{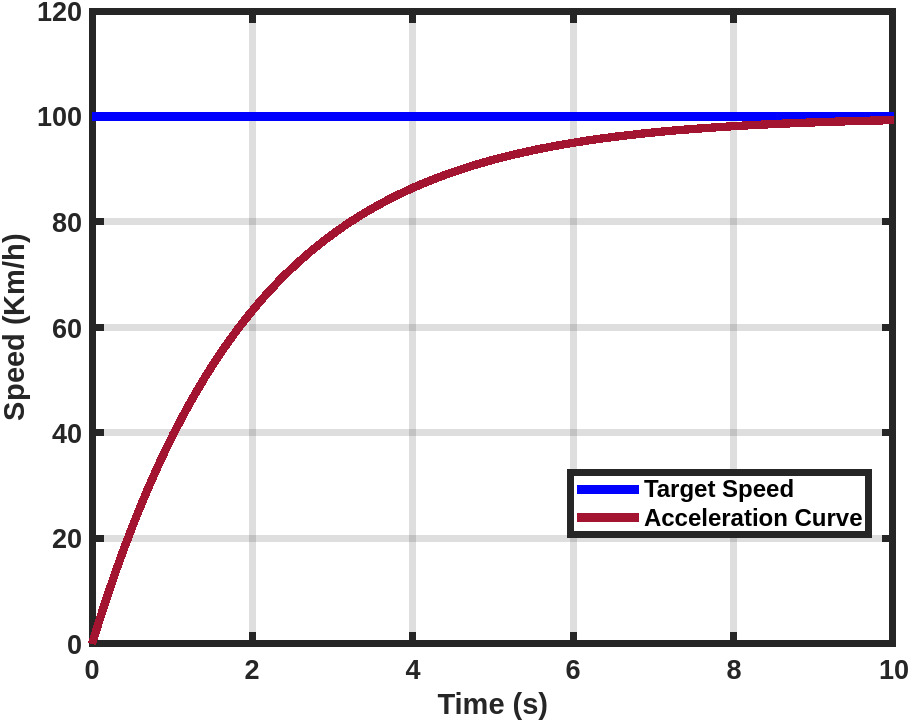}
    \caption{Acceleration Curve}
    \Description{Acceleration curve}
\end{minipage}
\hfill
\begin{minipage}[b]{0.48\textwidth}
    \centering
    \includegraphics[width=1.15\textwidth]{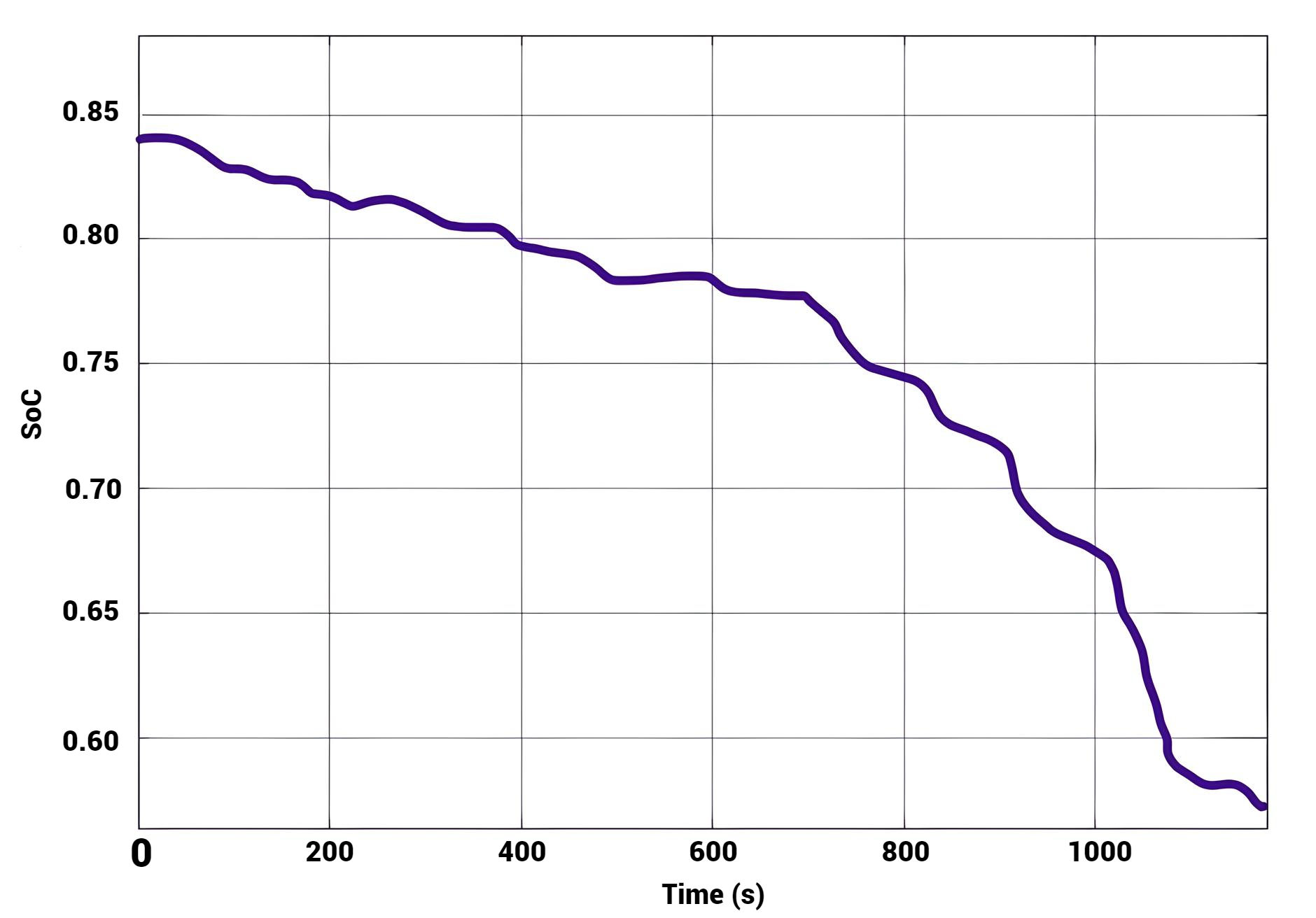}
    \caption{SoC Dynamics during UDDS Cycle}
    \Description{SoC Dynamic During UDDS Cycle}
\end{minipage}
\end{figure}

\begin{figure}[ht]
\centering
\includegraphics[width=0.6\textwidth]{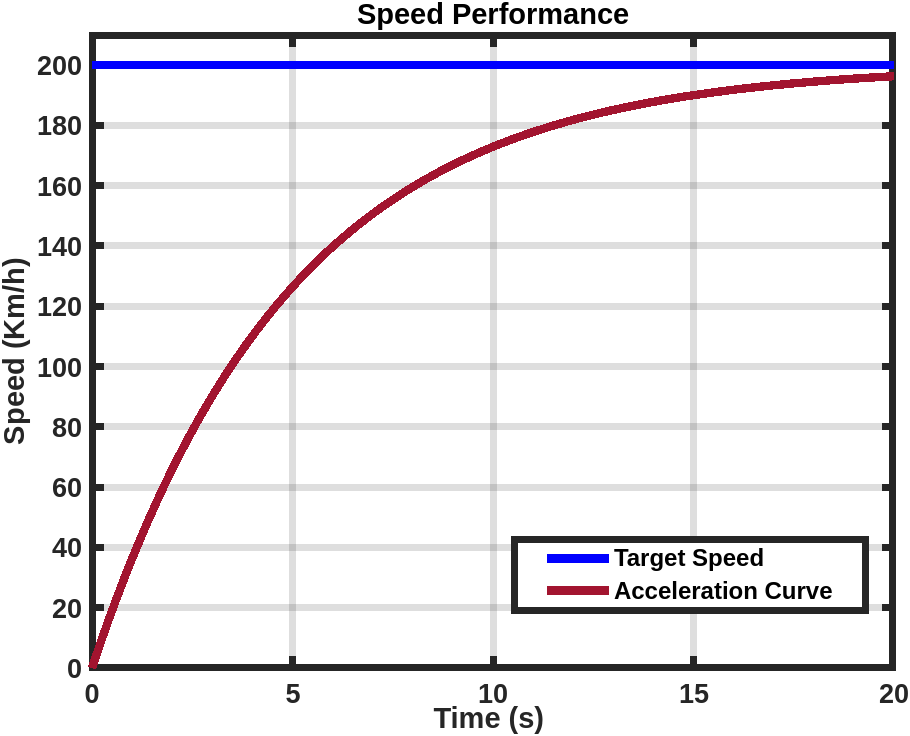}
\caption{Maximum Speed Curve}
\Description{Max Speed Curve}
\end{figure}

\subsection{\textbf{Battery SoC Dynamics}}
During the UDDS cycle, the variation in the battery's SoC is illustrated in Figure 11. The graph shows that at lower speeds, the AEV consumes less mechanical energy per unit time, resulting in a slower decrease in SoC. Conversely, higher speeds lead to a more rapid SoC decrease due to increased energy consumption, consistent with the principles of energy conservation. The energy recovery braking system, which transforms braking energy into stored electrical energy, is responsible for the periodic increase in SoC.

\subsection{\textbf{Top Speed Performance}}
In this paper, we set a target speed of 200 km/h to evaluate the AEV's top speed and recorded the maximum achievable speed, as illustrated in Figure 12. The autonomous EV reached a maximum speed of approximately 190 km/h, with an acceleration time from 0 to top speed of about 19 seconds. This indicates robust acceleration performance and a maximum speed that comfortably meets the needs of typical driving conditions.

\section{CONCLUSION}
Focusing on the critical but limited aspects of autonomous electric vehicle (AEV) research, this paper presents the development and detailed models for the driver, battery, motor, transmission, and vehicle body using MATLAB/Simulink. These models were assembled and integrated to create a pure AEV simulation. After configuring the necessary parameters, the simulation was conducted under UDDS standard cycle conditions to evaluate various performance metrics. The results demonstrate exceptional speed tracking accuracy, with the maximum deviation between actual and target speeds remaining under 1.5\%. In terms of travel distance, the AEV can cover over 350 km with a battery SoC reduction of 0.8, effectively meeting most users' daily driving requirements. The acceleration performance was also notable, with the AEV reaching 100 km/h in approximately 9.5 seconds. Additionally, the integration of an energy recovery braking system significantly enhanced efficiency, reducing energy consumption and extending the driving range by 25.5\% compared to vehicles without this system. Overall, the results show that the developed model offers robust performance, making it well-suited for real-world applications.

\section{LIMITATIONS, FUTURE RESEARCH \& TECHNICAL RECOMMENDATIONS}
\textbf{\textit{Limitation: Absence of Real-World Field Testing}}:
This study simulates real-time, real-world scenarios within the MATLAB environment but lacks actual field testing. While the simulations are designed to replicate the complexities and dynamics of real-world conditions, providing a controlled environment for initial modeling and analysis, the study’s findings remain unverified in practical, unpredictable scenarios that autonomous electric vehicles (AEVs) would face in real-life operations. This limitation underscores the need for future research to extend beyond simulated environments and incorporate field testing to fully validate the effectiveness and robustness of the proposed models and algorithms in real-world applications.

The following technical recommendations are suggested:

\begin{enumerate}
\item \textbf{Software Tools:}
\begin{itemize}
\item \textbf{MATLAB \& Simulink:} These are essential tools for developing Autonomous Electric Vehicles (AEVs). They offer a wide range of libraries and toolboxes to help design powertrains, understand vehicle dynamics, and model control systems.
\item \textbf{Simscape:} This is especially useful for creating detailed models of mechanical, electrical, and hydraulic systems. It helps make simulations more realistic and accurate.
\item \textbf{CarSim:} This software is important for simulating how a vehicle behaves under different driving conditions. It helps test and improve the algorithms that control autonomous driving.

\end{itemize}

\item \textbf{Modeling Approaches:}
\begin{itemize}
    \item \textbf{Vehicle Dynamics:} Creating a model that includes six degrees of freedom (6-DOF) is recommended. This helps simulate realistic vehicle behavior, such as how it steers and brakes, which is important for understanding how the vehicle will perform.
    \item \textbf{Electric Powertrain:} Detailed models of the electric motor, battery, and regenerative braking system are needed. This helps optimize how the vehicle uses energy, which affects its overall performance and efficiency.
    \item \textbf{Sensors and Actuators:} Including models for key sensors (like LiDAR and cameras) and actuators (like steering and braking systems) is crucial. This ensures the vehicle can accurately detect its surroundings and control its movements.
\end{itemize}

\item \textbf{Simulation Scenarios:}
\begin{itemize}
    \item \textbf{Diverse Conditions:} It’s important to simulate how the vehicle performs on different types of roads and in various weather conditions, such as rain or snow. This helps ensure that the vehicle's systems are robust and can handle real-world conditions.
    \item \textbf{ADAS and Autonomy:} Testing Advanced Driver Assistance Systems (ADAS) and autonomous driving algorithms using Hardware-in-the-Loop (HIL) simulations are recommended. HIL testing allows you to combine real hardware with simulation models for more realistic testing.
    \item \textbf{Energy Management:} Simulations should focus on managing energy use, especially through regenerative braking. This can help improve the vehicle’s range and efficiency.
\end{itemize}

\item \textbf{Validation and Testing:}
\begin{itemize}
    \item \textbf{HIL Testing:} Using Hardware-in-the-Loop (HIL) simulations is important for testing how the vehicle’s hardware and software work together in real time. This helps ensure that everything functions as expected under simulated conditions.
    \item \textbf{Real-World Trials:} To ensure that the vehicle performs well in real life, it’s important to do tests on actual tracks or using dynamometers. These tests verify that the simulation models are accurate and that the vehicle is ready for real-world use.
\end{itemize}
\end{enumerate}

\textbf{Acknowledgements:} We extend our gratitude to the University of Lagos for their library support.\\

\textbf{Declarations:} The authors have no conflicts of interest to declare that are relevant to the content of this article.\\

\textbf{Data Availability Statement:} All data analyzed during this study are included within the article.\\


\begin{thebibliography}{00}

\bibitem{b1} J. Wang and K. Zhang, "Research on modeling and simulation of a pure electric vehicle," in 2024 7th International Conference on Energy, Electrical and Power Engineering (CEEPE), IEEE, 2024. \url{https://doi.org/10.1109/CEEPE62022.2024.10586462}

\bibitem{b2} Q. Ajao and L. G. Sadeeq, "Overview analysis of recent developments on self-driving electric vehicles," arXiv preprint arXiv:2307.00016, 2023. \url{https://doi.org/10.48550/arXiv.2307.00016}

\bibitem{b3} K. T. Chau, "Pure electric vehicles," in Alternative Fuels and Advanced Vehicle Technologies for Improved Environmental Performance, Woodhead Publishing, 2014, pp. 655-684. \url{https://doi.org/10.1533/9780857097422.3.655}

\bibitem{b4}  A. Karki, B. L. Smith, R. Yadav, and S. Hossain, "Status of pure electric vehicle power train technology and prospects," Applied System Innovation, vol. 3, no. 3, pp. 35, 2020. \url{https://doi.org/10.3390/asi3030035}

\bibitem{b5} Z. Shen, J. Wang, A. Yildirim, and M. J. Khanna, "Optimization models for electric vehicle service operations: A literature review," Transportation Research Part B: Methodological, vol. 128, pp. 462-477, 2019. \url{https://doi.org/10.1016/j.trb.2019.08.006}

\bibitem{b6} Z. Li, A. Khajepour, and J. Song, "A comprehensive review of the key technologies for pure electric vehicles," Energy, vol. 182, pp. 824-839, 2019. \url{https://doi.org/10.1016/j.energy.2019.06.077}

\bibitem{b7}  I. Husain and M. S. Islam, "Design, modeling, and simulation of an electric vehicle system," SAE Transactions, pp. 2168-2176, 1999. \url{https://doi.org/10.4271/1999-01-1149}

\bibitem{b8} A. Koenig, K. H. Tok, S. Dörfler, and H. Herrmann, "Concept design optimization of autonomous and electric vehicles," in 2019 8th International Conference on Power Science and Engineering (ICPSE), IEEE, 2019. \url{https://doi.org/10.1109/ICPSE49633.2019.9041175}

\bibitem{b9} PerceptIn, "A self-driving car that guides you to the next destination," 2023.

\bibitem{b10} N. Ding, K. Prasad, and T. T. Lie, "The electric vehicle: a review," International Journal of Electric and Hybrid Vehicles, vol. 9, no. 1, pp. 49-66, 2017. Available: \url{https://doi.org/10.1504/IJEHV.2017.082816}

\bibitem{b11} N. Ding, K. Prasad, and T. T. Lie, "The electric vehicle: a review," International Journal of Electric and Hybrid Vehicles, vol. 9, no. 1, pp. 49-66, 2017. Available: \url{https://doi.org/10.1504/IJEHV.2017.082816}

\bibitem{b12} A. König, A. T. Dörfler, H. Herrmann, and S. Haas, "An overview of parameter and cost for battery electric vehicles," World Electric Vehicle Journal, vol. 12, no. 1, pp. 21, 2021. Available: \url{https://doi.org/10.3390/wevj12010021}

\bibitem{b13} A. O. Kıyaklı and H. Solmaz, "Modeling of an electric vehicle with MATLAB/Simulink," International Journal of Automotive Science and Technology, vol. 2, no. 4, pp. 9-15, 2019. Available: \url{https://doi.org/10.30939/ijastech..475477}

\bibitem{b14} Q. Ajao and L. Sadeeq, "Dynamic cell modeling of Li-Ion polymer batteries for precise SOC estimation in power-needy autonomous electric vehicles," arXiv preprint arXiv:2306.10654, 2023. Available: \url{https://doi.org/10.48550/arXiv.2306.10654}

\bibitem{b15} D. McDonald, "Electric vehicle drive simulation with MATLAB/Simulink," in Proceedings of the 2012 North-Central Section Conference, 2012.

\bibitem{b16} S. Kato, S. Tokunaga, Y. Maruyama, S. Maeda, M. Hirabayashi, Y. Kitsukawa, A. Monrroy, T. Ando, Y. Fujii, and T. Azumi, "Autoware on board: Enabling autonomous vehicles with embedded systems," in 2018 ACM/IEEE 9th International Conference on Cyber Physical Systems (ICCPS), 2018, pp. 287-296. Available: \url{https://doi.org/10.1109/ICCPS.2018.00035}

\end{thebibliography}
\end{document}